\documentstyle[aps,prd,preprint,psfig]{revtex}
\begin{document}
\tightenlines

\newcommand{\be}{\begin{equation}}
\newcommand{\ee}{\end{equation}}

\title{On the factor ordering problem in stochastic inflation}
\author{Alexander Vilenkin}

\address{Institute of Cosmology,
        Department of Physics and Astronomy,\\
        Tufts University,
        Medford, Massachusetts 02155, USA}
\date{\today}
\maketitle

\begin{abstract}

The stochastic approach to inflation suffers from ambiguities due to
the arbitrary choice of the time variable and due to the choice of the
factor ordering in the corresponding Fokker-Planck equation.  Here it
is shown that both ambiguities can be removed if we require that the
factor ordering should be set in such a way that physical results are
invariant with respect to time reparametrizations.  This requirement
uniquely selects the so-called Ito factor ordering.  Additional
ambiguities associated with non-trivial kinetic terms of the scalar
fields are also discussed, as well as ways of constraining these
ambiguities.

\end{abstract}

\section{Introduction.}

Quantum fluctuations of the inflaton field $\phi$ play an important
role in the inflationary scenarios.  These fluctuations can be
pictured as a random walk of $\phi$ superimposed on the deterministic
classical evolution and can be described in terms of a probability
distribution satisfying the Fokker-Planck (FP) equation \cite{1,2,3,4,5,6,7}.
This approach, however, suffers from significant ambiguities: the
resulting probabilities depend on the arbitrary choice of the time
variable $t$ and on the ordering of the non-commuting factors in the
FP equation \cite{6,7,8}.  

The origin of the dependence on the choice of $t$ has been extensively
discussed in the literature \cite{7,8,9,10}.  In essence, the problem is
that physical volumes of regions with the field $\phi$ in any given
range grow unboundedly with time.  The relative probabilities of
different values of $\phi$ are given by the corresponding volume
ratios, and we have the usual ambiguities arising when one tries to
compare infinities.  I suggested a possible resolution of this problem
in Ref.\cite{11} and will have nothing more to say about it here.

In this paper we shall concentrate on probability distributions for
which the divergence
of the physical volume is not relevant.  These include 
the coordinate (rather
than physical) volume distribution for $\phi$ and the physical volume
distribution in models where inflation is not eternal.  There are no 
infinities to deal with in these cases, and one might expect that
there will also be no
ambiguities.  One finds, however, that there is still some dependence
of the results on the choice of $t$, as well as on the factor
ordering.

It has been argued in Ref.\cite{8} that the factor ordering ambiguity
represents uncertainties inherent in the stochastic approach.  The
resulting uncertainties in the probabilities have been estimated as
$O(H^2)$, where $H$ is the inflationary expansion rate and I use
Planck units throughout the paper.  In models of new inflation with
$H\ll 1$ the uncertainties are small.  Moreover,
the uncertainties due to the choice of $t$ are also
$O(H^2)$, and it has been argued in \cite{8} that such uncertainties
are acceptable since they do not go beyond the factor ordering
ambiguity (which is presumably associated with the limitations of the
FP equation itself).

In the present paper I am going to take an alternative approach and
require that the factor ordering should be set in such a way that
appropriately chosen probability distributions are 
invariant with respect to time reparametrization.  I am going to show
that such an ordering does indeed exist (the so-called Ito ordering)
and suggest that it is the correct factor ordering to be used in the
FP equation.  This argument is presented in Section 3, after reviewing
the FP equation in Section 2.

In Section 4, I consider models with non-trivial kinetic terms for
$\phi$ which give rise to an additional factor-ordering ambiguity.  
I am going to argue that much of this ambiguity can be
removed by imposing some reasonable requirements on the form of the FP
equation.  The conclusions of the paper are briefly summarized 
in Section 5.

\section{The Fokker-Planck equation}

We shall consider a model with several inflaton fields $\phi^a$,
$a=1,...,N$, described by the Lagrangian
\begin{equation}
L={1\over{2}}\partial_\mu\phi^a\partial^\mu\phi^a - V(\phi).
\end{equation}
The evolution of $\phi^a$ during inflation is described by the 
probability distribution $P(\phi,t)d^N\phi$ which is interpreted, up
to a normalization, as the comoving volume of regions with specified
values of $\phi^a$ in the intervals $d\phi^a$ at time $t$.  The FP
equation for $P(\phi,t)$ has the form
\be
{\partial P\over{\partial t}}=-{\partial J^a\over{\partial\phi^a}},
\label{FP}
\ee
where the flux $J_a(\phi,t)$ is given by
\be
J_a=D(\phi)^{1/2+\beta}{\partial\over{\partial\phi^a}}[D(\phi)^{1/2-\beta}P]
-v_a(\phi)P.
\label{flux}
\ee
Here,
\be
D(\phi)=H(\phi)^{\alpha+2}/8\pi^2
\label{diffcoeff}
\ee
is the diffusion coefficient,
\be
H(\phi)=[8\pi V(\phi)/3]^{1/2}
\label{H}
\ee
is the inflationary expansion rate, and
\be
v_a(\phi)=-{1\over{4\pi
}}H(\phi)^{\alpha-1}{\partial\over{\partial\phi^a}}H(\phi)
\label{v}
\ee
is the ``drift'' velocity of the slow roll.  
$\phi$-space indices are raised and lowered using the flat metric
$\delta_{ab}$; hence, $J^a=J_a$, etc.

The parameter $\beta$ in Eq.(\ref{flux}) represents the ambiguity in
the ordering of the non-commuting factors $D(\phi)$ and
$\partial/\partial\phi^a$.  $\beta=0$ and $\beta=1/2$ correspond to
the so-called Stratonovich and Ito factor orderings, respectively.  
The parameter $\alpha$ in
Eqs.(\ref{diffcoeff}),(\ref{v}) represents the freedom of choosing the
time variable $t$ which is assumed to be related to the proper time of
comoving observers $\tau$ by
\be
dt=H(\phi)^{1-\alpha}d\tau.
\label{alpha}
\ee
Hence, $\alpha=1$ corresponds to the proper time parametrization
$t=\tau$ and $\alpha=0$ corresponds to using the logarithm of the
scale factor as a time variable.

The FP equation applies in the region of $\phi$-space enclosed by the
thermalization boundary $S_*$ where the conditions of slow roll
are violated,
\be
\left| {\partial H\over{\partial\phi^a}}(\phi_*)\right|\sim 2\pi
H(\phi_*).
\label{slowroll}
\ee
Here, $\phi_*\in S_*$.  The boundary condition on $P$ requires that
diffusion vanishes on $S_*$:
\be
n^a{\partial\over{\partial\phi^a}}[D(\phi)^{1/2-\beta}P]=0,
\label{bc}
\ee
where $n^a$ is the normal to $S_*$.  Since (\ref{slowroll}) is an
order-of-magnitude relation, the exact location of the surface $S_*$
depends on the choice of a constant of order 1.  Although this
introduces an ambiguity in the calculation of $P$, we note that
diffusion is small in the region dominated by the slow roll, and the
ambiguity in the choice of $S_*$ (which necessarily lies in that
region) does not significantly influence the solution of the FP
equation \cite{Planck}.

I have assumed for simplicity that we are dealing with inflation of
the ``new'' type.  In the case of ``chaotic'' inflation we would have
another boundary where $V(\phi)\approx 1$.  Quantum gravity effects
become important
at this Planck boundary, and it is not clear what kind of boundary
condition has to be imposed there.  

The FP equation for the physical volume distribution ${\tilde
P}(\phi,t)$ has the form 
\be
{\partial {\tilde P}\over{\partial t}}=-{\partial
{\tilde J}^a\over{\partial\phi^a}}+3H^\alpha {\tilde P},
\label{physFP}
\ee
where 
\be
{\tilde
J}_a=D(\phi)^{1/2+\beta}{\partial\over{\partial\phi^a}}[D(\phi)^{1/2-\beta}
{\tilde P}]
-v_a(\phi){\tilde P}.
\label{physflux}
\ee
and the last term in (\ref{physFP}) represents the growth of the
physical volume due to the expansion of the universe.  The
factor-ordering parameter $\beta$ could in principle be different for
the coordinate and physical volume distributions.

\section{Ito factor ordering and the time reparametrization invariance}

Let us now identify probability distributions that can be expected to
be invariant under time reparametrization.  The functions $P(\phi,t),
{\tilde P}(\phi,t)$
are not suitable for this role, since they give distributions for
$\phi$ on surfaces of constant $t$.  Dufferent choices of the time
variable $t$ result in different surfaces and different
distributions.  Mathematically, this is evident from the fact that the
corresponding FP equations explicitely depend on the
parameter $\alpha$.

Let us consider a large comoving region which is defined by a
spacelike hypersurface $\Sigma$ at some initial time $t=0$.  We
shall consider a family of time variables parametrized by $\alpha$ as
in Eq.(\ref{alpha}) assuming however that the surfaces $t=0$ coincide
with $\Sigma$
for all these variables.  Different parts of our comoving region will
have different evolution histories and will thermalize with different
values of $\phi\in S_*$.  At any time $t$ there will be parts of the
region that are still inflating, but in the limit $t\to\infty$ all the
comoving volume will be thermalized, except a part of measure zero.  

Let us introduce the distribution $p(\phi_*)dS_*$ which maps the
$\phi$-space thermalization boundary $S_*$ onto our comoving
volume.  It is defined as
the fraction of the comoving volume that is going to thermalize at
$\phi_*\in S_*$ in the surface element $dS_*$ (at any time).  The
distribution $p(\phi_*)$ is defined without reference to any
particular time variable, and we can expect it to be invariant with
respect to time reparametrization.  Using Eq. (\ref{flux}) for the
flux and the boundary condition (\ref{bc}) we can express $p(\phi_*)$
as
\be
p(\phi_*)=n^a(\phi_*)v_a(\phi_*)\int_0^\infty P(\phi_*,t)dt,
\label{0}
\ee
where $v_a(\phi)$ is given by Eq.(\ref{v}).  Now I am going to show
that with Ito factor ordering, $\beta=-1/2$, $p(\phi_*)$ is indeed
independent of the parameter $\alpha$.

Introducing
\be
\psi(\phi)=H^{\alpha-1}(\phi)\int_0^{\infty}P(\phi,t)dt
\label{psi}
\ee
and integrating the FP equation (\ref{FP}) and the boundary condition
(\ref{bc}) over $t$, we obtain
\be
{\partial\over{\partial\phi^a}}\left[{1\over{8\pi^2}}
{\partial\over{\partial\phi^a}}(H^3\psi)-{1\over{4\pi}}{\partial
H\over{\partial \phi^a}}\psi\right]=-P_0(\phi),
\label{1}
\ee
\be
n^a{\partial\over{\partial\phi^a}}(H^3\psi)=0 ~~~~~~~~~~~~~~~(\phi\in
S_*).
\label{2}
\ee
Here, $P_0(\phi)=P(\phi,0)$ is the initial distribution at $t=0$ and I
have used $\beta=-1/2$ in Eq.(\ref{1}).  The thermalization boundary
distribution $p(\phi_*)$ can also be expressed in terms of $\psi$,
\be
p(\phi_*)=-{1\over{4\pi}}{\partial
H\over{\partial\phi^a}}n^a\psi(\phi_*).
\label{3}
\ee

The function $\psi(\phi)$ is uniquely determined by Eq.(\ref{1}) with
the boundary condition (\ref{2}).  It can then be used in Eq.(\ref{3})
to evaluate the distribution $p(\phi_*)$.  Note that the parameter
$\alpha$ has been absorbed in the definition (\ref{psi}) of
$\psi(\phi)$ and does not appear in Eqs.(\ref{1})-(\ref{3}).  This
shows that $p(\phi_*)$ is independent of time parametrization.

It is not difficult to verify that the parameter $\alpha$ drops out of
the equations for $p(\phi_*)$ only in the case of $\beta=-1/2$
corresponding to Ito factor ordering.  We conclude, therefore, that
Ito ordering is the correct choice to be used in the FP equation \cite{J}.


Turning now to the physical volume FP equation (\ref{physFP}), we
consider models in which inflation is not eternal, that is, all
eigenvalues of the operator on the right hand side of
(\ref{physFP}) are negative.  Then, at $t\to\infty$ all physical
volume should be thermalized, except perhaps a part of measure zero.
We can define the physical volume distribution ${\tilde
p}(\phi_*)dS_*$, up to a normalization, as the physical volume of the
regions which thermalized with a given $\phi_*\in S_*$ in the surface
element $dS_*$.  The physical volume is measured at the time of
thermalization, so ${\tilde p}(\phi_*)$ is a mapping of $S_*$ onto the
thermalization hypersurface $\Sigma_*$ which separates the inflating
and thermalized regions of spacetime.  When inflation is not eternal,
the distribution ${\tilde p}(\phi_*)$ should be well defined and
independent of the time parametrization.  Now, it is easily seen that
Eqs.(\ref{0})-(\ref{3}) for the coordinate volume distribution can be
applied to ${\tilde p}(\phi_*)$ as well, with only a trivial
modification due to the last term in (\ref{physFP}).  The same
argument goes through, and we conclude again that ${\tilde p}(\phi_*)$
is invariant under time reparametrizations only with the choice of Ito
factor ordering, $\beta=-1/2$.  This argument cannot be directly
applied to models of eternal inflation, but it is natural to expect
that the same factor ordering will apply in such models as well.

\section{Diffusion on a $\phi$-space with a non-trivial metric}

Let us now consider models with non-trivial kinetic terms in the
scalar field Lagrangian,
\be
L={1\over{2}}K_{ab}(\phi)\partial_\mu\phi^a\partial^\mu\phi^b - V(\phi).
\ee
Such models arise, in particular, in the context of ``modular
inflation'' \cite{12}.
The matrix $K_{ab}(\phi)$ has the meaning of the metric on the space
of $\phi^a$ (the moduli space).  We shall assume that the expansion
rate $H(\phi)$ is small compared to the characteristic scale
$\Delta\phi$ on which 
$K_{ab}(\phi)$ vary in the $\phi$-space.  In modular inflation, we
expect $\Delta\phi\sim 1$, so this condition
is satisfied when the scale of inflation is well below the Planck
scale, $H(\phi)\ll 1$.

The fields $\phi^a$ can be thought of as coordinates in 
$\phi$-space, and the form of the FP equation should be invariant with
respect to the transformations
\be
\phi^a\to{\phi^a}'(\phi^b).
\ee
Coordinates can always be chosen so that $K_{ab}=\delta_{ab}$ at any
point in $\phi$-space.  We then expect 
that in the vicinity of that point the equation will take the form
(\ref{FP}),(\ref{flux}), possibly with corrections suppressed by some
powers of $H/\Delta\phi$.  These
conditions are satisfied by the simplest covariant generalization of
the flat-metric equation (\ref{FP}),(\ref{flux}),
\be
{\partial P\over{\partial t}} =
{1\over{\sqrt{K}}}{\partial\over{\partial\phi^a}} \left\{
\sqrt{K}K^{ab}\left[{\partial\over{\partial\phi^b}}(DP)-v_bP
\right]\right\}= \nabla_a[\nabla^a(DP)-v^aP] =\nabla_aJ^a.
\label{covFP}
\ee
Here, $K^{ab}(\phi)$ is the contravariant metric satisfying
\be
K^{ab}(\phi)K_{bc}(\phi)=\delta^a_c,
\ee
$K=\det (K_{ab})$, $\nabla_a$ is a covariant derivative with respect
to 
$\phi^a$, 
and I have used Ito factor ordering $\beta=-1/2$ in (\ref{flux}).
Note that the conservation of probability,
\be
{\partial\over{\partial t}}\int P\sqrt{K}d^N\phi = -\int_{S_*}J^adS_a,
\ee
follows immediately from (\ref{covFP}).

We note, however, that the conditions of covariance and correspondence
with the flat-metric form alone do not fix the form of the equation
uniquely.  One  could, for example, replace the covariant derivative in
(\ref{covFP}) by a more general expression
\be
\nabla^a\to\nabla_a +\xi H^2{\cal R}\nabla^a +\xi'H^2 {\cal
R}^{ab}\nabla_b +...,
\label{general}
\ee
where ${\cal R}(\phi)$ and ${\cal R}^{ab}(\phi)$ are respectively the
scalar curvature and the Ricci tensor of the $\phi$-space and
$\xi,\xi'$ are numerical coefficients.  The powers of $H$ in the
additional terms in (\ref{general}) are determined by dimensionality.
Since ${\cal R}\sim(\Delta\phi)^{-2}$, these terms are suppressed by a
factor $(H/\Delta\phi)^2$.  Hence, Eq.(\ref{covFP}) can be used as an
approximate FP equation in models with $H\ll\Delta\phi$.

\section{Summary and discussion}

The most satisfactory way to determine the factor ordering in the FP
equation would be to derive it from first principles.  At present we
do not have such a derivation.  The approach I took in this paper is
phenomenological: the factor ordering is determined by imposing some
physically reasonable requirements on the form of the FP equation.  I
have shown that for a scalar field model with a flat metric in the
$\phi$-space, $K_{ab}(\phi)=\delta_{ab}$, the factor ordering is
uniquely determined by requiring that physical results should not
depend on the arbitrary choice of the time variable.  This approach
selects the Ito factor ordering.

Additional ambiguities arising in models with a non-trivial metric
$K_{ab}(\phi)$ are constrained by requiring (i) covariance with
respect to coordinate transformations in the $\phi$-space and (ii)
correspondence with the flat metric case.  This fixes the form of the
FP equation up to factors of the order $(H/\Delta\phi)^2$ in the
diffusion term, where $\Delta\phi$ is the characteristic scale of
variation of $K_{ab}(\phi)$.


\section{Acknowlegements}

I am grateful to Serge Winitzki for useful discussions.  This work was
supported in part by the National Science Foundation.


\begin{references}

\bibitem{1}  A. Vilenkin, Phys. Rev. {\bf D27}, 2848 (1983).

\bibitem{2}  A. A. Starobinsky, in {\it Current topics in Field
Theory, Quantum Gravity and Strings}, edited by H. J. de Vega and N. Sanchez
(Springer, Heidelberg, 1986).

\bibitem{3}  A. S. Goncharov, A. D. Linde, and V. F. Mukhanov, Int. J.
Mod. Phys. {\bf A2}, 561 (1987).

\bibitem{4}  M. Mijic, Phys. Rev. {\bf D42}, 2469 (1990).

\bibitem{5}  Y. Nambu and M. Sasaki, Phys. Lett. {\bf B219}, 240
(1989); Y. Nambu, Prog. Theor. Phys. {\bf 81}, 1037 (1989); K. Nakao, Y.
Nambu, and M. Sasaki, Prog. Theor. Phys. {\bf 80}, 1041 (1988).

\bibitem{6}  D. S. Salopek and Bond, Phys. Rev. {\bf 43}, 1005 (1991).

\bibitem{7}  A. D. Linde, and A. Mezhlumian, Phys. Lett. {\bf B307},
25\ (1993); A. D. Linde, D. A. Linde, and A. Mezhlumian, Phys. Rev. {\bf D49}%
, 1783 (1994).

\bibitem{8} S. Winitzki and A. Vilenkin, Phys. Rev. {\bf D53}, 4298
(1996).

\bibitem{9}  A. Vilenkin, Phys. Rev. {\bf D52}, 3365 (1995).

\bibitem{10}  A. D. Linde and A. Mezhlumian, Phys. Rev. {\bf D53}, 4267
(1996).

\bibitem{11} A. Vilenkin, Phys. Rev. Lett. {\bf 58}, 5501 (1998).

\bibitem{Planck} Linde and Mezhlumian \cite{7} used the absorbing
boundary condition, $P=0$, at the Planck boundary.  We note that with
this boundary condition, the argument of Section III goes through
without change.

\bibitem{J} A somewhat related argument suggesting that Ito factor
ordering is the correct choice was given in J. Garriga and
A. Vilenkin, Phys. Rev. {\bf D57}, 2230 (1998), Section V(D).

\bibitem{12} See, e.g., T. Banks {\it et. al.}, Phys. Rev. {\bf D52},
3548 (1995).

\end{references}
\end{document}